\documentclass[aps,prd,twocolumn,showpacs,groupedaddress,floatfix]{revtex4-1}

\usepackage{graphicx}
\usepackage{array}
\usepackage{amsmath}
\usepackage{amssymb}
\usepackage{amsfonts}

\DeclareGraphicsExtensions{.eps, .eps, .jpg, .png}

\def\e{{\mathrm{e}}}

\newcommand{\de}{\mathrm{d}}

\begin{document}

\title{Tachyonic Quintessence and a Preferred Direction in the Sky}

\author{Luca Grisa}
\email[]{lgrisa@physics.umass.edu}
\affiliation{Department of Physics, University of Massachusetts, Amherst, MA, 01003}

\author{Lorenzo Sorbo}
\email[]{sorbo@physics.umass.edu}
\affiliation{Department of Physics, University of Massachusetts, Amherst, MA, 01003}

\date{\today}

\begin{abstract}

We show that quintessence, when it is described by a tachyonic field, can amplify a tiny primordial gradient generating a preferred direction in the sky. In its simplest realization, this mechanism only affects the Cosmic Microwave Background fluctuations at the quadrupole level. We briefly discuss how higher multipoles can also be affected, once the full structure of the quintessence potential is taken into account.

\end{abstract}

\maketitle

%%%%%%%%%%%%%%%%%%%%%%%%
\section{Introduction}%%
%%%%%%%%%%%%%%%%%%%%%%%%

The formulation of the cosmological principle~\cite{Einstein:1917ce} coincides with the birth of modern scientific cosmology. Over the last decades, homogeneity and isotropy of the Universe have been tested to increasing degrees of accuracy. In particular, the radiation of the Cosmic Microwave Background (CMB) is one of the most sensitive probes to test the isotropy of the Universe. The observed temperature of the CMB is observed to be uniform at zeroth order (once the dipole is subtracted): this result led to the formulation of the theory of primordial inflation. Moreover, approximate statistical isotropy appears to hold even for the small fluctuations of the CMB temperature.

In spite of these striking results, several analyses of the CMB fluctuation maps, starting from~\cite{Eriksen:2003db,Hansen:2004vq} (see, {\em e.g.}, \cite{Hanson:2009gu} for a more recent analysis), have shown the existence of anomalies associated to some degree of breaking of statistical isotropy. Even though a clear consensus on the subject is still absent ({\em e.g.}, see \cite{Bennett:2010jb} for an up to date discussion) intensive work has been done in the past also as a response to the theoretical challenge of naturally generating a preferred directions in the sky. The models proposed usually rely either on early-Universe or on late-Universe mechanisms. While the former lead to an {\em intrinsic} anisotropy in the primordial spectrum of metric perturbations, the latter produce anisotropy in the {\em observed} CMB via anisotropic Sachs-Wolfe effect.  Early-Universe mechanisms include~\cite{Ackerman:2007nb} (see however~\cite{Himmetoglu:2008hx}) and~\cite{Watanabe:2009ct}, that rely on vector fields or~\cite{Boehmer:2007ut}, that rely on spinor fields, as well as~\cite{Donoghue:2007ze,Erickcek:2008sm}, that rely on primordial gradients. Late-Universe mechanisms can invoke magnetic fields~\cite{Campanelli:2007qn} or anisotropies in the dark energy equation of state~\cite{Rodrigues:2007ny,Koivisto:2008ig,Battye:2009ze}. Despite those many attempts, it is fair to say that breaking statistical isotropy usually requires strong and aesthetically unappealing assumptions.

In the present work we show a (relatively) natural mechanism that can give rise to a preferred direction in the CMB sky. This mechanism is a hybrid of early and a late Universe ones and does not rely on vector fields or on unusual properties of the dark energy sector, but rather on a simple model of scalar dark energy, characterized by a non-trivial, tachyonic ($V''(\phi)<0$) potential.

Models of tachyonic quintessence frequently appear in the literature. For instance, if the quintessential scalar is described by a pseudo-Nambu-Goldstone Boson (pNGB)~\cite{Frieman:1995pm}, then half of the extrema of the potential $V(\phi)\propto 1+\cos(\phi/f)$ are tachyonic. Another scenario of tachyonic quintessence was proposed in~\cite{Kallosh:2002gf}, where the magnitude of the tachyonic mass is related to the height of the potential at its maximum.

Our main assumption is that a tachyonic field $\phi$ has a small primordial gradient. As the Hubble parameter drops below $\sqrt{\left|V''(\phi)\right|}$, the initial gradient is amplified by the ``pull'' of its tachyonic potential, effectively converting a small primordial isocurvature perturbation into a larger curvature mode. In first approximation, this leads to an ``ellipsoidal'' Universe~\cite{Campanelli:2007qn} which could explain the low quadrupole CMB amplitude observed in both COBE and WMAP data~\cite{Contaldi:2003zv}, if the preferred direction in the quintessence correlated with that of the primordial quadrupole.

As we will see, our mechanism will be at work for a tachyonic mass a few times larger than $\Lambda/M_\mathrm{Pl}^2$. This is, in particular, the case for a pNGB with value of the axion constant $f$ slightly sub-Planckian.

The primordial gradient in $\phi$ can be a remnant of the epoch that preceded the last bout of inflation, provided that such last bout were sufficiently short (this argument will be discussed in greater detail in section~\ref{gradient}). In this respect, our scenario could shed some light on pre-inflationary dynamics, similarly to the situations studied, {\em e.g.}, in~\cite{Chang:2007eq} and especially~\cite{Donoghue:2007ze}. In our case, a constant gradient can be the dominant remnant of a more general inhomogeneous initial value of the tachyonic field, {\em i.e.}, we can expand $\phi\approx \kappa_0+\kappa_1\,z_p+\kappa_2\,z^2_p+\dots$ -- for sake of simplicity, we will only consider powers of the $z$ coordinate, though the argument for the dominance of the linear term would still hold, were more generic terms considered -- at $N_\mathrm{i}\equiv N_{\mathrm{obs}}+\delta N$ e-foldings before the end of inflation ($N_{\mathrm{obs}}$ corresponds to the time when the current cosmological scales left the horizon and $z_p$ to the physical distance along the $z$ direction). We assume that each term $\kappa_k\,z_p^k$ had given an equal contribution to the energy in $\phi$ at that moment. Today, the scales, that left the horizon $N_{\mathrm{obs}}+\delta N$ e-foldings before the end of inflation, are still outside the horizon by a factor of $\e^{\delta N}$, {\em i.e.}, they correspond to physical distances of the order of $\e^{\delta N}\,H_0^{-1}$. Assuming no (other) significant evolution in $\phi$~\footnote{The evolution in $\phi$ occurs only at sub-horizon scales and in the late Universe, when the Hubble parameter is smaller than the (absolute value of the tachyonic) mass of quintessence, and it is not sufficient to invalidate this argument.}, the assumption that each term $\kappa_k\,z_p^k$ be of the same order implies that $\kappa_k\approx \e^{-k\,\delta N}\,H_0^k$. As a consequence, inside the horizon $z_p\approx H_0^{-1}$, each term $\kappa_k\,z_p^k$ contributes like $\e^{-k\,\delta N}$ to the energy in $\phi$. This implies that the terms with the lowest powers in $z_p$ will give the largest contributions to the metric perturbations at sub-horizon scales. For this reason, our analysis will be focused on the approximation where, in the early Universe, $\phi=\kappa_1\,z$ with $\phi(z=0)$ being set equal to zero by an appropriate choice of the origin of the $z$ coordinate.\\

The paper is organized as follows. In section~\ref{metric} we set up our system and solve the corresponding Einstein equations. Sections~\ref{rs} and~\ref{cmb} contain the effect of our anisotropic metric on the redshift and in particular on the CMB quadrupole anisotropy. Section~\ref{gradient} contains a discussion of the magnitude of the primordial gradient responsible for the anisotropy. Finally, in the concluding section we will discuss how our mechanism could also lead to alignments of the higher multipoles.

%%%%%%%%%%%%%%%%%%%%%%%%%%%%%%%%%%%%
\section{The metric}\label{metric}%%
%%%%%%%%%%%%%%%%%%%%%%%%%%%%%%%%%%%%

We consider the cylindrically symmetric cosmological metric
\begin{equation}
\de s^2=-\de t^2+a(t,\,z)^2\,\left(\de x^2+\de y^2\right)+b(t,\,z)^2\,\de z^2\,\,.
\end{equation}
The non-vanishing components of the Einstein tensor are
\begin{eqnarray}
G_{tt}&=&\frac{2}{b^2}\,\frac{a'}{a}\,\frac{b'}{b}-\frac{1}{b^2}\,\left(\frac{a'{}^2}{a^2}+2\,\frac{a''}{a}\right)+\frac{\dot{a}^2}{a^2}+2\,\frac{\dot{a}}{a}\,\frac{\dot{b}}{b},\nonumber\\
G_{xx}&=&G_{yy}=a^2\,\left(-\frac{1}{b^2}\,\frac{a'}{a}\,\frac{b'}{b}+\frac{1}{b^2}\frac{a''}{a}-\frac{\dot{a}}{a}\frac{\dot{b}}{b}-\frac{\ddot{a}}{a}-\frac{\ddot{b}}{b}\right)\,\,,\nonumber\\
G_{tz}&=&2\,\frac{a'}{a}\frac{\dot{b}}{b}-2\frac{\dot{a}'}{a}\,\,,\nonumber\\
G_{zz}&=&b^2\left(\frac{1}{b^2}\frac{a'{}^2}{a^2}-\frac{\dot{a}^2}{a^2}-2\frac{\ddot{a}}{a}\right)\,\,,
\end{eqnarray}
where a dot denotes a derivative with respect to $t$ and a prime a derivative with respect to $z$.

We assume the presence of three contributions to the stress-energy tensor: pressure-less dust with energy density $\rho$, a cosmological constant $\Lambda$ and a scalar field $\phi$ with canonically normalized kinetic term and potential $V(\phi)$ -- in principle one can always consider $\Lambda$ to be a part of $V(\phi)$ however we prefer to keep these two components separate as we will keep separate the cosmological constant problem and the breaking of statistical isotropy.

The components of the stress energy tensor of $\phi$ read
\begin{eqnarray}
T_{tt}&=&\frac{1}{2}\,\dot{\phi}^2+\frac{1}{2\,b^2}\,\phi'{}^2+V(\phi)+\Lambda\,\,\nonumber\\
T_{xx}&=&T_{yy}=a^2\left[\frac{1}{2}\dot{\phi}^2-\frac{1}{2\,b^2}\,\phi'{}^2-V(\phi)-\Lambda\right]\,\,,\nonumber\\
T_{zt}&=&\dot\phi\,\phi'\,\,,\nonumber\\
T_{zz}&=&b^2\left[\frac{1}{2}\dot{\phi}^2+\frac{1}{2\,b^2}\,\phi'{}^2-V(\phi)-\Lambda\right]\,\,,
\end{eqnarray}
and throughout the paper we will consider $\phi$ to be a tachyon with potential $V(\phi)=-m^2\,\phi^2/2$. The stress energy tensor for  dust has the general form $T_{\mu\nu}=\rho\,u_\mu\,u_\nu$, with $u^\mu\,\nabla_\mu u^\nu=0$.

Let us now solve the above equations by assuming that the non-isotropic, non-homogeneous scalar field is the source of a perturbation around the Friedmann-Robertson-Walker metric of a $\Lambda\mathrm{CDM}$ Universe. As a consequence, we write $a(t,\,z)=a_0(t)\,[1+\delta a(t,\,z)]$ and $b(t,\,z)=a_0(t)\,[1+\delta b(t,\,z)]$, where $\delta a\ll 1$ and $\delta b\ll 1$; the Friedmann equation for the background $\phi=0$ has exact solution 
\begin{equation}\label{scalefactor}
a_0(t)=\left(\frac{1-\Omega_\Lambda}{\Omega_\Lambda}\right)^{1/3}\,\sinh^{2/3}\left(\frac{3}{2}\,\sqrt{\Omega_\Lambda}\,H_0\,t\right)\,\,,
\end{equation}
where $H_0$ is the current value of the Hubble parameter and $\Omega_\Lambda\sim0.3$ from observation. 

We can assume that the dust velocity field $u^\mu$ is equal to $(1+\delta u^0,\, 0,\,0,\,\delta u^z)$ with $\delta u^0$ and $\delta u^z$ being first order quantities. Thus the geodesic equation for $u^\mu$ reduces to $\delta\dot{u}^0=0$ and $\left(a_0^2\,\delta u^z\right)^.=0$. Choosing $\delta u^z=0$ as $t\rightarrow 0$, we obtain, up to second order in the perturbations, $u^\mu=(1,\,0,\,0,\,0)$. Then conservation of the stress-energy tensor for dust implies that $\rho=\rho_0/(a^2\,b)$, where $\rho_0$ is the dust density at a fixed time.

Turning on the scalar field $\phi$, we assume that, as described in the introduction (see also~\cite{Turner:1991dn,Langlois:1995ca,Erickcek:2008sm}), the lowest terms in the expansion in powers of the coordinate $z$ give the largest contributions to the anisotropy. Then, without loss of generality, we can set $\phi(z=0)=0$ (remember that $\phi=0$ corresponds to the maximum of the tachyonic potential), so that
\begin{equation}
\phi(t,\,z)=\kappa_1(t)\,z+{\cal {O}}(z^2)\,\,.
\end{equation}
At leading order, the Klein-Gordon equation $\ddot\phi+3H\dot\phi-\phi''/a_0^2-m^2\,\phi=0$ decomposes as
\begin{eqnarray}
\ddot\kappa_1+3\frac{\dot a_0}{a_0}\,\dot\kappa_1-m^2\,\kappa_1=0\,\,.\label{k1}
\end{eqnarray}

We then expand $\delta a(t,\,z)=\delta a_0(t)+\delta a_2(t)\,z^2+{\cal {O}}(z^3)$ and $\delta b(t,\,z)=\delta b_0(t)+\delta b_2(t)\,z^2+{\cal {O}}(z^3)$ (it is easy to see that, since at this level of approximation $\phi$ is odd in $z$, $\delta a$ and $\delta b$ must be even in $z$). The $(tz)$ Einstein equation -- at first order $-2\,\delta \dot{a}'=\phi'\,\dot\phi/M_\mathrm{Pl}^2$ -- gives rise to 
\begin{eqnarray}
\delta\dot{a}_2&=&-\frac{\kappa_1\,\dot\kappa_1}{4\,M_\mathrm{Pl}^2}\,\,.\label{deltaa2}
\end{eqnarray}

The $(zz)$ Einstein equation reads
\begin{equation}
-2\,\delta\ddot{a}-6\,\frac{\dot{a}_0}{a_0}\,\delta\dot a=\frac{1}{M_\mathrm{Pl}^2}\left(\frac{1}{2}\dot\phi^2+\frac{1}{2\,a_0^2}\phi'{}^2+\frac{m^2}{2}\phi^2\right)\,\,.
\end{equation}
The only non-trivial part of this equation is the one independent on $z$, as the rest is redundant with respect to the two previous equations:
\begin{equation}\label{deltaa0}
-2\,\delta\ddot{a}_0-6\,\frac{\dot{a}_0}{a_0}\,\delta\dot a_0=\frac{\kappa_1^2}{2\,a_0^2\,M_\mathrm{Pl}^2}\,\,.
\end{equation}

The $(tt)$ Einstein equation reads, at first order in $\delta a$ and $\delta b$ and using the background equations,
\begin{eqnarray}
&&-2\frac{\delta a''}{a_0^2}+2\,\frac{\dot a_0}{a_0}\,\left(2\,\delta \dot a+\delta \dot b\right)=\\
&&=\frac{1}{M_\mathrm{Pl}^2}\left[\frac{1}{2}\dot\phi^2+\frac{1}{2\,a_0^2}\phi'{}^2-\frac{m^2}{2}\phi^2-\frac{\rho_0}{a_0^3}\left(2\,\delta a+\delta b\right)\right]\,\,,\nonumber
\end{eqnarray}
where the only term under control in the $z^n$ expansion is the one in $z^0$ (since $\delta a$ is determined with $\mathcal{O}(z^3)$ accuracy, the presence of the term $\delta a''$ constrains the accuracy of this equation to be $\mathcal{O}(z)$ hence, at this level of approximation, $\delta b$ can be determined only in its $z^0$ component). Therefore the only equation we obtain from the $(tt)$ component is 
\begin{eqnarray}
\label{equdeltab0}
-4\frac{\delta a_2}{a_0^2}+2\,\frac{\dot a_0}{a_0}\,\delta \dot \psi_0&=&\frac{1}{M_\mathrm{Pl}^2}\left(\frac{\kappa_1^2}{2\,a_0^2}-\frac{\rho_0}{a_0^3}\,\delta \psi_0\right)\,\,,
\end{eqnarray}
where $\delta\psi_0\equiv 2\,\delta a_0+\delta b_0$.\\

The explicit expressions of $\kappa_1(t)$, $\delta a_0(t)$ and $\delta b_0(t)$ are the following. The solution of eq.~\eqref{k1}, using the expression~\eqref{scalefactor} for $a_0(t)$, reads
\begin{equation}
\kappa_1(t)=\frac{\bar\kappa_1\,H_0\,M_\mathrm{Pl}}{\sqrt{1+\mu^2}}\frac{\sinh(\sqrt{1+\mu^2}\tau)}{\sinh\tau}
\end{equation}
where we defined the dimensionless quantities $\tau\equiv\frac{3}{2}\sqrt{\Omega_\Lambda}H_0\,t$ and $\mu^2\equiv4\,m^2/(9\,\Omega_\Lambda H_0^2)$, and where the integration constant $\bar\kappa_1$ is defined so that $\kappa_1(0)=\bar\kappa_1\,H_0\,M_\mathrm{Pl}$.

It is straightforward to solve for $\delta a_2$ from eq.~\eqref{deltaa2} using the found solution for $\phi(t,\,z)$ at the desired order:
\begin{eqnarray}
\delta a_2(t)&=&-\frac{\kappa_1(t)^2-\bar\kappa_1^2H_0^2M_\mathrm{Pl}^2}{8\,M_\mathrm{Pl}^2}\,\,,
\end{eqnarray}
where we have imposed $\delta a_2(t\rightarrow 0)=0$.

$\delta a_0$ can be computed by integrating eq.~\eqref{deltaa0}:
\begin{eqnarray}
\delta a_0(t)=-\frac{1}{9\,\Omega_\Lambda H_0^2M_\mathrm{Pl}^2}\left(\frac{\Omega_\Lambda}{1-\Omega_\Lambda}\right)^{2/3}\times\nonumber\\
	\times\int_0^\tau\frac{\de\tau_1}{\sinh^2\tau_1}
	\int_0^{\tau_1}\kappa_1^2(\tau_2)\,\sinh^{2/3}\tau_2\,\de\tau_2
\end{eqnarray}
where once again $\delta a_0(t\rightarrow 0)=0$.

Lastly, we compute the correction to the scale factor along the $z$ direction, $\delta b_0$, from \eqref{equdeltab0}:
\begin{eqnarray}
\delta b_0(t)=-2\,\delta a_0(t)+\frac{\bar\kappa_1^2}{10\,\Omega_\Lambda}\left(\frac{\Omega_\Lambda}{1-\Omega_\Lambda}\right)^{2/3}\times\nonumber\\
	\times\phantom{.}_2F_1\left(\frac{3}{2},\frac{5}{6},\frac{11}{6},-\sinh^2\tau\right)\cosh\tau\sinh^{2/3}\tau\,\,.
\end{eqnarray}

%%%%%%%%%%%%%%%%%%%%%%%%%%%%%%%%%%%%%%%%%%%%
\section{The perturbed redshift}\label{rs}%%
%%%%%%%%%%%%%%%%%%%%%%%%%%%%%%%%%%%%%%%%%%%%

An observable quantity related to the metric found in the previous section is the angular dependence of the CMB radiation induced by the integrated Sachs-Wolfe effect on our system. We compute it in this and in the following section. Photons coming from the last scattering surface move along null geodesics:
\begin{equation}
\frac{\de k^\mu}{\de\lambda}+\Gamma^\mu_{\nu\rho}\,k^\nu\,k^\rho=0\,\,,
\end{equation}
where $\lambda$ is the affine parameter along the worldline $x^\mu(\lambda)$ of the photon. $k^\mu=\de x^\mu/\de\lambda$ denotes the tangent (null) vector to the photon worldline.

It is convenient to introduce the conformal time $\eta$, so that the metric reads
\begin{eqnarray}
\de s^2=&&a_0^2(\eta)\,\left[-\de\eta^2+(1+2\,\delta a(\eta,\,z))\,\left(\de x^2+\de y^2\right)+\right.\nonumber\\
&&\left.+(1+2\,\delta b(\eta,\,z))\,\de z^2\right]\,\,.
\end{eqnarray}
This metric will allow us to factor out the dependence of the redshift $\zeta$~\footnote{We denote the redshift by $\zeta$ because $z$ is already chosen for one of the coordinates.} on the scale factor so to effectively simplify the calculation. Let us perform the conformal transformation
\begin{equation}\label{conformal}
\de{s}^2=a_0^2(\eta)\,\de\bar{s}^2\,,\ \de{\lambda}=a_0^2(\eta)\,\de{\bar\lambda}\,,\ {k}^\mu=a_0(\eta)^{-2}\bar{k}^\mu\,\,.
\end{equation}
It is easy to show by taking advantage of $k_\mu$ being a null vector, that $\bar k^\mu$ satisfies
\begin{equation}
\frac{\de\bar k^\mu}{\de\bar\lambda}+\bar\Gamma^\mu_{\nu\rho}\,\bar k^\nu\,\bar k^\rho=0\,\,,
\end{equation}
where $\bar\Gamma^\mu_{\nu\rho}$ are the Christoffels symbols of the metric $\de\bar s^2$, which is a first order perturbation around the Minkowski metric.

The photon worldline, $x^\alpha(\bar\lambda)$, can be expanded in a perturbation series
\begin{equation}
x^\mu(\bar\lambda)=x_B^\mu(\bar\lambda)+\delta x^\mu(\bar\lambda)\,,\ \bar{k}^\mu(\bar\lambda)=\bar{k}_B^\mu(\bar\lambda)+\delta\bar{k}^\mu(\bar\lambda)\,\,,
\end{equation}
where the subscript $B$ denotes a background quantity and $\delta$ denotes the first-order perturbation on it.

Since the background component of $\de\bar s^2$ describes a Minkowski spacetime, then the background null vector $\bar{k}_B^\mu$ is constant. By appropriately scaling $\bar\lambda$ we impose $\bar k_B^0=-1$. We have $\bar{k}_B^\mu=x_B^\mu{}'(\bar\lambda)=(-1,\,\vec{n})$, where $\vec{n}$ is a unit 3-vector, as we will eventually choose $\bar\lambda=0$ for the observer of the CMB photon, with $\lambda$ increasing as we go back in time. Given the symmetries of our spacetime, we can locate  the observer at $(x_B)_O=(y_B)_O=0$, while $(z_B)_O=z_0$ remains arbitrary, so we  have
\begin{eqnarray}\label{flatsol}
&&\eta_B(\bar\lambda)=\eta_0-\bar\lambda\,\,,\\
&&x_B(\bar\lambda)=n_x\,\bar\lambda\,,\ y_B(\bar\lambda)=n_y\,\bar\lambda\,,\ z_B(\bar\lambda)=z_0+n_z\,\bar\lambda\,\,.\nonumber
\end{eqnarray}

The redshift of a source with four-velocity $u^\mu_S$, as seen by an observer whose four-velocity is $u^\mu_O$, is defined as
\begin{equation}\label{redshift}
1+\zeta=\frac{{({g}_{\mu\nu} {k}^\mu {u}^\nu)}_{S}} 
{{({g}_{\alpha\beta} {k}^\alpha {u}^\beta)}_{O}}\,\,.
\end{equation}
It is convenient to introduce a ``conformal'' redshift $\bar\zeta$, defined  by adding a bar to all the quantities in eq.~\eqref{redshift} and defining $u^\mu=a_0^{-1}\,(\eta) \bar{u}^\mu$:
\begin{equation}\label{confredsh}
1+\zeta=\frac{1+\bar\zeta} {a_0(\eta_S)}\,\,,
\end{equation}
where we have used the fact that $a_0(\eta_0)=1$. We assume that the spatial components of $\bar u^\mu_S$ and $\bar u^\mu_O$ can be treated as first order quantities, in which case $\bar\zeta$ is also a first order quantity. By using $\bar{u}^\mu_{O,\,S}=(1,\,\vec{v}_{O,\,S})$, where $\vec{v}_O$ is the peculiar velocity of the observer and ${\vec{v}}_S$ of the source, we compute $\bar\zeta$ to be
\begin{equation}
\bar\zeta\approx\left({\vec{v}}_S-{\vec{v}}_O\right)\cdot\vec{n}- \delta \bar{k}^0(\bar\lambda_S)+\delta \bar{k}^0(\bar\lambda=0)\,\,,
\end{equation}
where we have used the fact that the photon was emitted at $\bar\lambda=\bar\lambda_S$ and is observed at $\bar\lambda=0$. The quantity $\delta \bar{k}^0({\bar{\lambda}_S})-\delta \bar{k}^0(0)$ is obtained by integrating the geodesic equation for $\bar k^0$
\begin{equation}
\frac{\de\,\delta\bar k^0}{\de\bar\lambda}+\frac{\partial\,\delta a}{\partial\eta}\left(n_x^2+n_y^2\right)+\frac{\partial\,\delta b}{\partial\eta}\,n_z^2=0\,\,.
\end{equation}

Therefore $\bar\zeta$ is
\begin{equation}\label{finalz}
\bar\zeta\approx\left(\vec{v}_S-\vec{v}_O\right)\cdot\vec{n}+
\int_0^{\bar{\lambda}_S}\left[\frac{\partial\,\delta a}{\partial\eta}\left(n_x^2+n_y^2\right)+\frac{\partial\,\delta b}{\partial\eta}\,n_z^2\right]\,\de\bar\lambda\,\,,
\end{equation}
where the first term corresponds to the intrinsic motions of the source and of the observer. The second term is associated to the integrated Sachs-Wolfe effect on our spacetime; the next section will be dedicated to a thorough discussion of this term.

%%%%%%%%%%%%%%%%%%%%%%%%%%%%%%%%%%%%%%%
\section{CMB anisotropies}\label{cmb}%%
%%%%%%%%%%%%%%%%%%%%%%%%%%%%%%%%%%%%%%%

The effect of our anisotropic space on the CMB spectrum is given by the integrated Sachs-Wolfe (iSW) effect:
\begin{equation}\label{zeta}
\bar\zeta_{\mathrm{iSW}}\approx \int_0^{\bar\lambda_S}\left[\frac{\partial\,\delta {a}}{\partial\eta}\,(n_x^2+n_y^2)+\frac{\partial\,\delta {b}}{\partial\eta}\,n_z^2\right]\,\de\bar{\lambda}\,\,.
\end{equation}

Using~\eqref{flatsol}, we trade $\bar\lambda$ for $\eta$, moreover, since $\delta a$ and $\delta b$ rapidly vanish as $\eta\rightarrow 0$, we can extend the integration to $\eta=0$ ({\em i.e.}, $\bar\lambda_S=\eta_0$). We remind that $\delta a\approx\delta a_0(\eta)+\delta a_2(\eta)\,z^2$ and $\delta b\approx\delta b_0(\eta)$. Since $\delta a_2$ and $\delta b_2$ appear on the same footing in the equations above, and since $\delta b_2$ is undetermined at this level of approximation, we neglect also $\delta a_2$ in what follows. Thus we have
\begin{equation}
\bar\zeta_{\mathrm{iSW}}\approx \int_0^{\eta_0}\left(\frac{\de\,\delta a_0}{\de\eta}\sin^2\vartheta+\frac{\de\,\delta b_0}{\de\eta}\cos^2\vartheta\right)\,\de\eta\,\,,
\end{equation}
where $n_z\equiv\cos\vartheta$.

The above expression can be integrated explicitly
\begin{equation}
\bar\zeta_{\mathrm{iSW}}\approx \delta a_0(\eta_0)+\left(\delta b_0(\eta_0)-\delta a_0(\eta_0)\right)\,\cos^2\vartheta\,\,,
\end{equation}
in which we have used $\delta a_0(\eta=0)=\delta b_0(\eta=0)=0$. It is assumed that the direction of alignment of the gradient coincides precisely with the $z$ axis. In order to find the equivalent expression when the gradient of $\phi$ is directed along a direction $(\Theta,\,\Phi)$, we rotate the coordinate system $(x,\,y,\,z)$ by an angle $\Phi$ around the $z$ axis and by an angle $\Theta$ around the $x$ axis. We thus obtain
\begin{eqnarray}
\bar\zeta_{\mathrm{iSW}}(\vartheta,\,\varphi)&&\approx \delta a_0(\eta_0)+\left(\delta b_0(\eta_0)-\delta a_0(\eta_0)\right)\times\\
&&\times\left[\sin\vartheta\,\sin\Theta\,\cos(\Phi+\varphi)+\cos\vartheta\,\cos\Theta\right]^2\,\,.\nonumber
\end{eqnarray}

We follow, {\em e.g.}, \cite{Koivisto:2008ig} to find the effect on CMB. If $T_*$ is the CMB temperature at decoupling, then the observed temperature will be $T(\vartheta,\,\varphi)=T_*/(1+\zeta(\vartheta,\,\varphi))\simeq T_*\,a_{\mathrm{CMB}}\,(1-\bar\zeta(\vartheta,\,\varphi))$, where $a_{\mathrm{CMB}}$ is the scale factor at decoupling. Then the average observed temperature is $\langle T\rangle=\int\de\varphi\,\de\cos\vartheta\,T(\vartheta,\,\varphi)/4\pi$ and the anisotropy $\delta T(\vartheta,\,\varphi)/T=1-T(\vartheta,\,\varphi)/\bar{T}=\bar\zeta_{\mathrm{iSW}}(\vartheta,\,\varphi)-\int\de\varphi\,\de\cos\vartheta\,\bar\zeta_{\mathrm{iSW}}(\vartheta,\,\varphi)/4\pi$. Its decomposition in spherical harmonics is given by
\begin{equation}
a_{\ell m}=\int\de\varphi\,\de\cos\vartheta\frac{\delta T(\vartheta,\,\varphi)}{T}\,Y_\ell^m{}^*\,\,.
\end{equation}

An explicit calculation allows to show that the only non-vanishing contributions from the anisotropic integrated Sachs-Wolfe effect are
\begin{eqnarray}
a^{\mathrm{iSW}}_{22}&=&\sqrt{\frac{2\,\pi}{15}}\,\left(\delta b_0(\eta_0)-\delta a_0(\eta_0)\right)\,\sin^2\Theta\,\e^{2i\Phi}\,\,,\\
a^{\mathrm{iSW}}_{21}&=&-\sqrt{\frac{2\,\pi}{15}}\,\left(\delta b_0(\eta_0)-\delta a_0(\eta_0)\right)\,\sin\,2\Theta\,\e^{i\Phi}\,\,,\nonumber\\
a^{\mathrm{iSW}}_{20}&=&\frac{1}{3}\,\sqrt{\frac{\pi}{5}}\,\left(\delta b_0(\eta_0)-\delta a_0(\eta_0)\right)\,\left(1+3\,\cos\,2\Theta\right)\,\,,\nonumber
\end{eqnarray}
with $a_{2,-1}=-a_{21}^*$ and $a_{2,-2}=a_{22}^*$.

These contributions have to be summed to the intrinsic quadrupole momentum of the CMB. It is possible to see that our contribution can help explaining the low quadrupole observed in CMB data, since it effectively gives rise to an ``ellipsoidal'' Universe~\cite{Campanelli:2007qn}.

Let us quickly review the argument of~\cite{Campanelli:2007qn}: we denote by $a_{2m}^I$  (with $a^I_{2,-m}=(-1)^m\,a^I_{2m}{}^*$) the quadrupole components of the ``intrinsic'' CMB fluctuations. As a consequence of statistical isotropy, the $a^I_{2m}$ can be assumed to be equal up to a phase, {\em i.e.}, $a^I_{20}=\sqrt{\pi/3}\,{\cal Q}_I$, $a^I_{21}=\sqrt{\pi/3}\,\e^{i\,\alpha_1}\,{\cal Q}_I$, $a^I_{22}=\sqrt{\pi/3}\,\e^{i\,\alpha_2}\,{\cal Q}_I$.

The quadrupole amplitude is defined by $Q_2^2=\frac{3}{5\,\pi}\sum_m\left|a^{\mathrm{iSW}}_{2m}+a_{2m}^I\right|^2$. If $\Theta$, $\Phi$ and $\left(\delta b_0(\eta_0)-\delta a_0(\eta_0)\right)$ are appropriately chosen, then one can obtain a ``low'' value of $Q_2\simeq 5.3\times 10^{-6}$ even if the prediction for a spherical Universe, ${\cal {Q}}_I\simeq 1.3\times 10^{-5}$, would be higher. In particular, from the analysis of~\cite{Campanelli:2006vb}, we can extract the required value of $\left|\delta b_0(\eta_0)-\delta a_0(\eta_0)\right|\simeq 2\times 10^{-5}$.

\begin{figure}[]
   \begin{center}
   \includegraphics[width=3in]{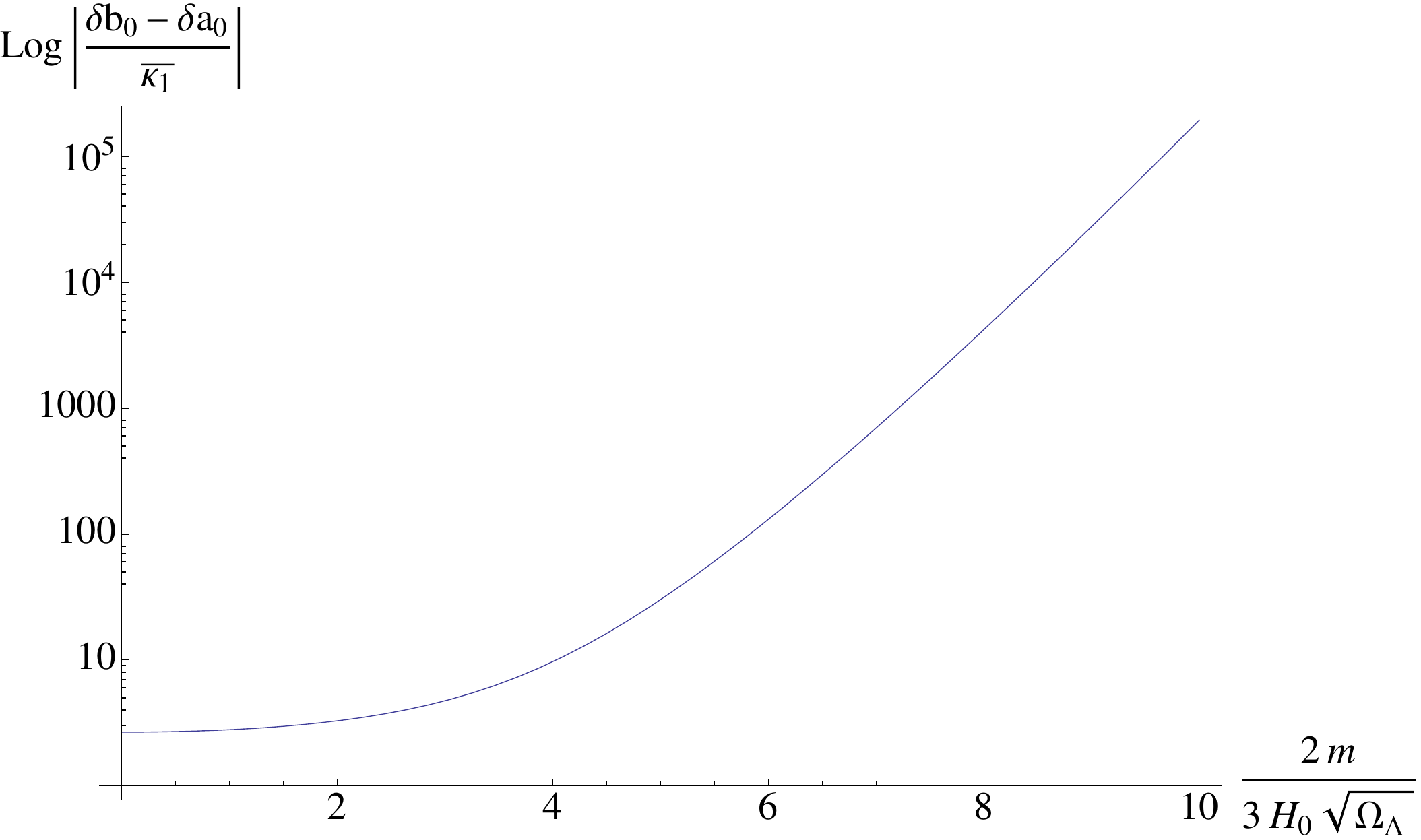}
   \caption{The quantity $|\delta b_0(\eta_0)-\delta a_0(\eta_0)|/\bar\kappa_1^2$ plotted as a function of $\mu$.}
   \label{fig1}
   \end{center}
\end{figure}

It is however important to note that the cancellation of the intrinsic quadrupole component by the integrated Sachs-Wolfe effect associated to the ellipsoidal Universe requires a correlation between the primordial and the late components, as discussed already in~\cite{Contaldi:2003zv} and quantified in~\cite{Gruppuso:2007ya,Appleby:2009za}. Even in the absence of correlations with the intrinsic quadrupole, the effect on the CMB provides the strongest constraints on our scenario: those constraints can be expressed as $\left|\delta b_0(\eta_0)-\delta a_0(\eta_0)\right|\lesssim 2\times 10^{-5}$. In figure~\ref{fig1}, we show the log-plot of $\left|\delta b_0(\eta_0)-\delta a_0(\eta_0)\right|/\bar\kappa_1^2$ as a function of the parameter $\mu$. This plot shows how, for $m\gtrsim {\cal {O}}(\mathrm{few})\times H_0$, the degree of anisotropy grows exponentially with $m$. Note also that, even for $m=0$, we can have a small anisotropy imprinted on the CMB. As we will discuss below, however, in this case the original gradient in $\phi$ would not have been negligible during inflation, and its effect should have been taken into account in the calculation of the primordial spectrum of density fluctuations.

%%%%%%%%%%%%%%%%%%%%%%%%%%%%
\section{On the initial gradient}\label{gradient}%%
%%%%%%%%%%%%%%%%%%%%%%%%%%%%

In this section we discuss the magnitude of the initial gradient $\kappa_1(0)\equiv \bar\kappa_1\,H_0\,M_\mathrm{Pl}$. Since for $H\gg m$ the evolution of the field $\phi$ is frozen by Hubble friction, we have $\phi\simeq \kappa_1(0)\,z$ inside our Hubble patch. As a consequence, the energy $\rho_\phi=T_{tt}^\phi$ scales as $\kappa_1(0)^2/\left(2\,a_0(t)^2\right)$. This gradient term redshifts as $a(t)^{-2}$, {\em i.e.}, as spatial curvature. During inflation, with Hubble parameter $H_I$, $\phi$ provides a fraction
\begin{equation}
\frac{\rho_\phi}{\rho_{\mathrm{tot}}}\simeq \frac{\bar\kappa_1^2}{6}\,\frac{H_0^2}{H_I^2}\,\frac{\e^{2\,N}}{a_{\mathrm{end}}^2}
\end{equation}
of the background inflaton energy, where $N$ is the number of e-foldings from the end of inflation and $a_{\mathrm{end}}$ is the scale factor at the end of inflation. In particular, if inflation lasted $N_{\mathrm{i}}$ e-foldings and we denote by $T_{RH}\simeq (3\,M_\mathrm{Pl}^2\,H_I^2)^{1/4}$ the reheating temperature (we assume instantaneous reheating and ignore the effects of $g_*$, the effective number of relativistic degrees of freedom in the system), then we can trade $\kappa_1(0)$ for $(\rho_\phi/\rho_{\mathrm{tot}})|_\mathrm{i}$ computed at the beginning of inflation
\begin{equation}
\bar\kappa_1\simeq \sqrt{2\left.\frac{\rho_\phi}{\rho_\mathrm{tot}}\right|_\mathrm{i}}\,\frac{T_0\,T_{RH}}{H_0\,M_\mathrm{Pl}}\,\e^{-N_{\mathrm{i}}}\,\,,
\end{equation}
where we have used $a_{\mathrm{end}}=T_0/T_{RH}$ with $T_0\simeq 2\times 10^{-4}$eV being the current value of the CMB temperature.

As discussed in section~\ref{cmb}, data require $|\delta b_0(\eta_0-\delta a_0(\eta_0)|\lesssim 2\times 10^{-5}$, that implies 
\begin{equation}
\left.\frac{\rho_\phi}{\rho_\mathrm{tot}}\right|_\mathrm{i}\lesssim 10^{-5}\,\left[\frac{\left|\delta b_0(\eta_0)-\delta a_0(\eta_0)\right|}{\bar\kappa_1^2}\right]^{-1}\,\frac{H_0^2\,M_\mathrm{Pl}^2}{T_0^2\,T_{RH}^2}\,\e^{2\,N_\mathrm{i}}\,\,,
\end{equation}
where the quantity $\left|\delta b_0(\eta_0)-\delta a_0(\eta_0)\right|/\bar\kappa_1^2$ is plotted in figure~\ref{fig1}.

Finally, we use the relation $\e^{N_{\mathrm{obs}}}\simeq (H_I\,T_0)/(H_0\,T_{RH})$ ($N_{\mathrm{obs}}$ is the number of observable e-foldings of inflation) to obtain the condition
\begin{equation}
\left.\frac{\rho_\phi}{\rho_\mathrm{tot}}\right|_\mathrm{i}\simeq 3\times 10^{-6}\,\left[\frac{\left|\delta b_0(\eta_0)-\delta a_0(\eta_0)\right|}{\bar\kappa_1^2}\right]^{-1}\,\e^{2\,(N_{\mathrm{i}}-N_{\mathrm{obs}})}\,\,.
\end{equation}

Therefore inflation can start with a sizable (but still smaller than unity) value of $(\rho_\phi/\rho_\mathrm{tot})|_\mathrm{i}$ provided $N_\mathrm{i}$ is sufficiently larger than $N_{\mathrm{obs}}$. Then when the observable scales left the horizon, $N_{\mathrm{obs}}$ e-foldings before the end of inflation, the perturbation in the scalar $\phi$ was a smaller fraction (by a factor $\exp\left[2\,(N_{\mathrm{obs}}-N_i)\right]$) of the background energy, small enough to affect neither the dynamics of the inflaton nor that of the perturbations. The gradient in $\phi$ stays irrelevant until the Hubble parameter is comparable with $m$, when it will start increasing under the effect of the tachyonic potential and will give a perturbation of the right amplitude today.

To fix ideas we can set $N_{\mathrm{obs}}=60$, $\mu=10$ (so that $\left|\delta b_0(\eta_0)-\delta a_0(\eta_0)\right|/\bar\kappa_1^2\simeq 10^5$), $N_\mathrm{i}=71$. Then, $(\rho_\phi/\rho_\mathrm{tot})|_\mathrm{i}\simeq 0.1$. At the time observable scales exited the horizon, $\rho_\phi/\rho_{\mathrm{tot}}$ would have redshifted by a factor $\e^{2\times (60-71)}\simeq 10^{-10}$ and it would have been negligible. In the absence of the tachyonic mass, the gradient of $\phi$ would be still negligible today. However, due to the tachyonic enhancement $\left|\delta b_0(\eta_0)-\delta a_0(\eta_0)\right|$, the effects of such a primordial gradient will manifest themselves in the lowest CMB multipoles~\footnote{The possibility of a gradient in the dark energy is also considered in~\cite{Gordon:2005ai}, where, however, it is assumed that the quintessence is so light that its dynamics is frozen.}.

Before concluding this section, let us discuss the possible worry that, since $\phi$ is effectively a massless field for most of the history of the Universe, its quantum fluctuations generated during inflation may be larger than the classical gradient that is central to our analysis. It is easy to choose the inflationary parameters in such a way that the quantum fluctuations are subdominant with respect to the original gradient. Indeed, the amplitude of the quantum fluctuations in $\phi$ will be of the order of $H_I$, the Hubble parameter during  inflation. During inflation, the modulation of $\phi$ at scales $z\simeq H_0^{-1}$, which are relevant today, is of the order of $\kappa_1(0)\,z\simeq \bar\kappa_1\,M_\mathrm{Pl}\simeq \sqrt{(\rho_\phi/\rho_\mathrm{tot})|_\mathrm{i}}\,\e^{N_{\mathrm{obs}}-N_\mathrm{i}}\,M_\mathrm{Pl}$. Assuming $(\rho_\phi/\rho_\mathrm{tot})|_\mathrm{i}={\cal {O}}(1)$, we see that, as long as $H_I\ll M_\mathrm{Pl}\,\e^{N_{\mathrm{obs}}-N_\mathrm{i}}$ ({\em e.g.}, $H_I\ll 10^{-5}\,M_\mathrm{Pl}$ in the example $N_\mathrm{i}-N_{\mathrm{obs}}=11$ considered above), the effects of quantum fluctuations are negligible.

%%%%%%%%%%%%%%%%%%%%%%
\section{Conclusions and Discussion}%%
%%%%%%%%%%%%%%%%%%%%%%

If inflation lasted a relatively short time, then some primordial gradients could exist as a relic of the chaotic pre-inflationary dynamics. A gradient in the quintessence field, even if too small to leave any detectable effects during the observable epoch of inflation, might be amplified by a tachyonic quintessence to have observable magnitude today. We have seen that in the simplest scenario, the dominant effect of the amplified gradient is on the CMB quadrupole, whose observed amplitude sets the strongest constraints on the parameters of the problem.

The fact that CMB fluctuations are affected only at the quadrupole level is a consequence of the approximation $\phi\simeq \kappa_1(t)\,z$. Higher powers of $z$ in the expansion of $\phi$ would lead to the generation of higher multipole contribution. Generically, terms of order $z^n$ in the expansion of the scalar field $\phi(t,z)$ will generate contributions on both $\delta a(t,z)$ and $\delta b(t,z)$ up to order $z^{n-1}$, which, in turn, provide terms up to order $z^{n+1}$ in $\bar\zeta_\mathrm{iSW}$ as for equation~\eqref{zeta}. That is, terms of order $z^n$ in $\phi(t,z)$ will affect $C_\ell$'s of $\ell=n+1$. Given the hints of multipole alignments up to $\ell\sim 40$, it would be interesting to be able to extend our mechanism  in such a way that larger powers of $z$ in the expression of $\phi(t,\,z)$ are generated. One possibility is that the self-interactions of $\phi$ during the recent cosmological evolution are responsible. For instance, if the quintessence field $\phi$ is given by a pseudo-Nambu-Goldstone boson, its potential is $V(\phi)=\Lambda\,\left[1+\cos\left(\phi/f\right)\right]/2$ ($f^2=\Lambda/(2\,m^2)$). By following our analysis of section~\ref{metric}, we see that, because of the Klein-Gordon equation, the coefficients of lower powers in $z$ will act as sources for those of higher powers, hence generating higher multipoles. A further investigation of this mechanism is subject of future work.\\

{\bf Acknowledgments.} This work has been supported in part by the NSF grant PHY - 0855119. We thank John Donoghue, Burak Himmetoglu, David Langlois, Nemanja Kaloper and Marco Peloso for interesting discussions.

\bibliographystyle{apsrev}
\bibliography{AnisoGS10}

\end{document}